\documentclass[%
 reprint,
superscriptaddress,
 amsmath,amssymb,
 aps,
]{revtex4-2}
\usepackage{amsmath}
\usepackage{graphicx}
\usepackage{dcolumn}
\usepackage{bm}
\usepackage{ar}
\usepackage{xcolor}
\usepackage{units}

\graphicspath{{./Figures/}}

\begin{document}

\title[Dimensional analysis of spring-wing systems]{Dimensional analysis of spring-wing systems reveals performance metrics for resonant flapping-wing flight }


\affiliation{%
 Mechanical and Aerospace Engineering,  University of California, San Diego
}%
\affiliation{%
 School of Physics,  Georgia Institute of Technology
}%

\author{James Lynch}
\affiliation{%
 Mechanical and Aerospace Engineering,  University of California, San Diego
}%
\author{Jeffrey Gau}
\affiliation{%
 School of Physics,  Georgia Institute of Technology
}%
\author{Simon Sponberg}
\affiliation{%
 School of Physics,  Georgia Institute of Technology
}%
\author{Nick Gravish}
 \email{To whom correspondence should be sent: ngravish@ucsd.edu}
\affiliation{%
 Mechanical and Aerospace Engineering,  University of California, San Diego
}%

\date{\today}

\begin{abstract}
Flapping-wing insects, birds, and robots are thought to offset the high power cost of oscillatory wing motion by using elastic elements for energy storage and return. Insects possess highly resilient elastic regions in their flight anatomy that may enable high dynamic efficiency. However, recent experiments highlight losses due to damping in the insect thorax that could reduce the benefit of those elastic elements. We performed experiments on, and simulations of a dynamically-scaled robophysical flapping model with an elastic element and biologically-relevant structural damping to elucidate the roles of body mechanics, aerodynamics, and actuation in spring-wing energetics. We measured oscillatory flapping wing dynamics and energetics subject to a range of actuation parameters, system inertia, and spring elasticity. To generalize these results, we derive the non-dimensional spring-wing equation of motion and present variables that describe the resonance properties of flapping systems: $N$, a measure of the relative influence of inertia and aerodynamics, and $\hat{K}$, the reduced stiffness. We show that internal damping scales with $N$, revealing that dynamic efficiency monotonically decreases with increasing $N$. Based on these results, we introduce a general framework for understanding the roles of internal damping, aerodynamic and inertial forces, and elastic structures within all spring-wing systems.

\end{abstract}

\maketitle

\section{Introduction}

\begin{figure*}[t]
    \begin{centering}
        \includegraphics[width=0.8\linewidth]{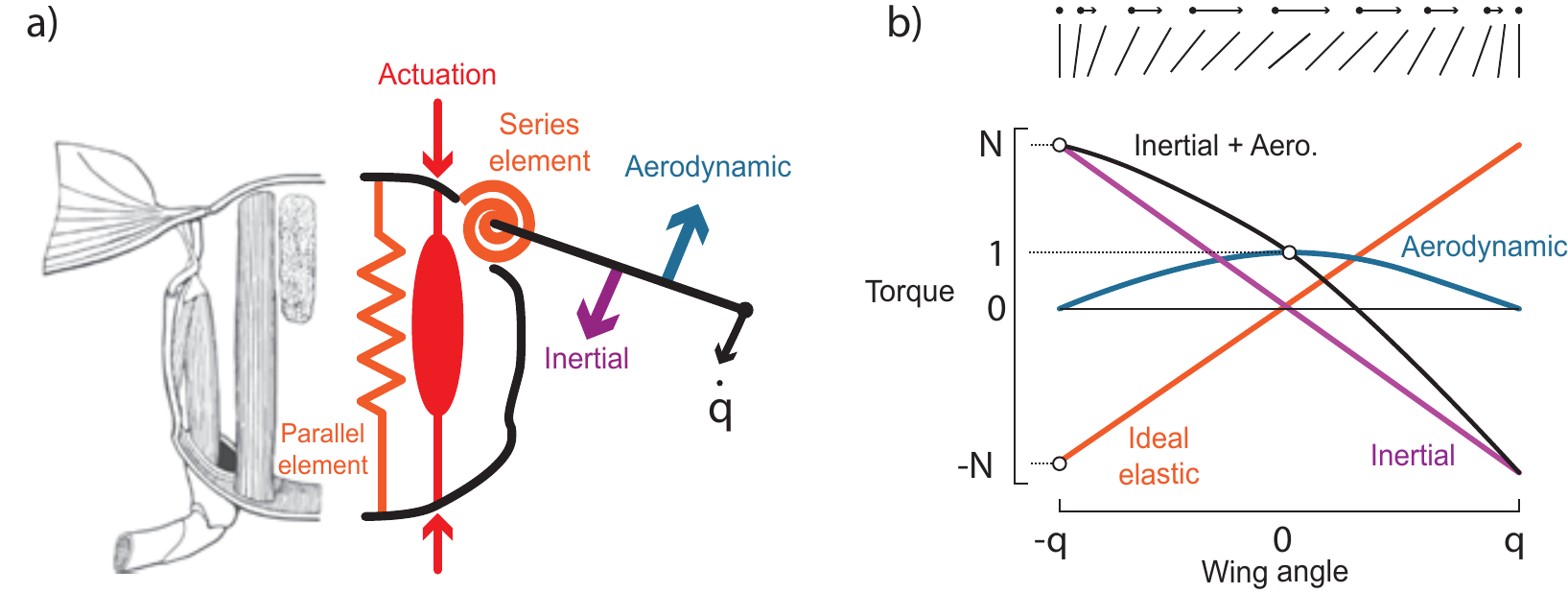}
        \caption{a) Flapping wing insects have elastic components in their thorax and wing hinge which can potentially act as spring elements to reduce the energetic demands of flapping wing flight. The left side of the image in a) shows an illustration of the muscle, thorax wall, and wing hinge of an insect. The right side of the image interprets it as a spring in parallel with the actuation source and a spring in series with the actuator, with aerodynamic and inertial forces that act on the rotating wing. b) Weis-fogh introduced a convenient way to visualize the relative magnitudes of torque acting on the wing hinge by normalizing all values to the peak aerodynamic torque. The units of both axes are dimensionless. The downstroke time of the insect in panel (a) is shown from the vertical dashed line. Image in (a) adapted from \cite{Snodgrass1935-db}.}
        \label{fig:spring_wing_intro}
    \end{centering}
\end{figure*}

Flapping wing flight is one of the most energetically demanding modes of locomotion in nature and in engineered flying robotic systems. Actuators must provide power to overcome aerodynamic forces on the wings, generate inertial forces for oscillatory acceleration and deceleration, and counteract internal energy losses from imperfect power transmission \cite{gau_indirect_2019}. If an oscillating wing is coupled to an elastic element such as a spring, the kinetic energy from the wing could be stored as elastic energy at the end of the wing-stroke and returned after stroke reversal. Many insects \cite{weis-fogh_rubber-like_1960,jensen_biology_1962,weis-fogh_energetics_1972,gau_indirect_2019, george_cross-bridge_2013}, birds \cite{ingersoll_how_2018,weis-fogh_energetics_1972}, and even bats \cite{konow_spring_2015} have spring-like elements in the form of elastic materials in their thoraxes, muscles, and tendons that may aid in reducing the energetic demands of flapping flight and improving flight efficiency (resonance) (Fig. \ref{fig:spring_wing_intro}a). However, the evidence that insects and birds operate near resonance largely relies on corelational observations of wingbeat frequency and wing inertia \cite{sotavalta_flight-tone_1952, greenewalt_wings_1960}, or energetics arguments comparing metabolic and aerodynamic power \cite{weis-fogh_energetics_1972, weis-fogh_quick_1973, dickinson_muscle_1995}. 

If animals do rely on elastic energy storage for improved efficiency, then there are implications for the dynamics and energetics of those systems. One major example is that, to benefit from a spring, flapping wings must be actuated at a specific resonance frequency governed by the spring stiffness, body morphology, and other factors such as aerodynamics and damping. Flapping at a higher or lower frequency leads to inevitable reduction in flight performance. Early experiments on the relationship between wingbeat frequency and wing inertia provided compelling evidence that insects do oscillate their wings at resonance \cite{sotavalta_flight-tone_1952,greenewalt_wings_1960}. However, these experiments relied only on the manipulation of wing inertia (without accompanying measurements of thorax stiffness) and thus do not provide direct comparison of the spring-wing system's resonant frequency and the wingbeat frequency. Recently, there has been some effort to measure insect thorax elasticity and frequency response \cite{jankauski_measuring_2020}, but experimental limitations leave room for questions about whether insects are, in fact, flapping at a resonant frequency. 

The impact of operation on or off resonance is directly related to how much a particular flapping wing system benefits from the inclusion of a spring. A flyer with small wing inertia would have less excess kinetic energy to store and return in a spring than one with large wing inertia, but also would be less impacted by changes in wingbeat frequency. The question of where insects fall on this spectrum was first address by Weis-Fogh in his analysis of flapping flight efficiency \cite{weis-fogh_quick_1973}. He introduced the dynamic efficiency ($\eta$), a ratio of aerodynamic work, $W_{aero}$, to actuator work, $W_{tot}$ that serves as a measure of how much energy is expended on useful aerodynamic work versus wasteful inertial work. Weis-Fogh provided an analysis of the scaling of $\eta$ in the absence of elasticity by introducing the non-dimensional variable $N$, which is the ratio of peak inertial force to peak aerodynamic force over a wing-stroke. In flapping-wing flight without springs, $\eta$ was shown to monotonically decrease with increasing $N$, thus requiring larger energy input at larger $N$ to sustain flight. In subsequent text we refer to $N$ as the Weis-Fogh number to reflect his contributions to flight energetics.

Examination of the sinusoidal motion of a wing in the absence of elasticity, reveals two sources of reaction force: inertial and aerodynamic forces. During a half-stroke of a wing, the inertial force associated with wing acceleration is at a maximum when the wing position is at its point of reversal, and inertial force decreases linearly with wing position and reaches zero at mid-stroke (Fig. \ref{fig:spring_wing_intro}b). At reversal, the wing speed is zero and thus the aerodynamic force at this point is zero, while at mid-stroke the aerodynamic force is maximum. Plotting these forces as a function of wing position (Fig. \ref{fig:spring_wing_intro}b) and normalized by peak aerodynamic force reveals that the inertial force has a maximum value of $N$, the Weis-Fogh number. Furthermore, the integration of these forces over the wing displacement provides the total inertial and aerodynamic work. As can be seen in Figure \ref{fig:spring_wing_intro}b an ideal spring exactly matched to the inertial force of the wing would exactly cancel out the inertial work over a cycle in a parallel spring-wing system. In such a case the dynamic efficiency of the system would be $\eta = 1$ and the system would be operating at resonance. However, it is less clear how internal damping, frequency modulation, and different spring arrangements modulate the dynamic efficiency.

The primary mechanism of elastic energy storage in insects is resilin, a highly-resilient elastic material first identified in patches of the locust exoskeleton was discovered and characterized in the 1960's by Weis-Fogh \cite{weis-fogh_rubber-like_1960,weis-fogh_energetics_1972,jensen_biology_1962} (and subsequently identified in many other arthropods). It was shown to return greater than 97\% of stored elastic energy, suggesting that insects have resilient components within their thorax that can facilitate efficient energy exchange and return. Thus, the historical choice not to include internal losses in the computation of dynamic efficiency appears to be a simplification based on the assumption that the losses due to aerodynamic forces are significantly larger.

However, recent experiments to characterize the energy storage and return in the Hawkmoth (\textit{Manduca sexta}) thorax have also demonstrated that small but significant energy loss occurs from internal structural damping \cite{gau_indirect_2019}. Similar structural damping was also observed in cyclic oscillation of cockroach legs \cite{Dudek2006-rz}, possibly suggesting a general character of energy loss in exoskeleton deformations. Structural damping is a form of energy loss different from the more familiar velocity-dependent viscous damping. Materials that are structurally damped exhibit energy loss that is frequency independent and is instead governed by oscillation amplitude and elastic coefficient \cite{beards_engineering_1995}. This is consistent with the interpretation of structural damping as a result of internal, microscopic friction that is not dependent on velocity. While the presence of highly elastic resilin suggests significant potential benefits from elastic materials, internal damping may preclude the energetic benefits of elasticity in spring-wing systems.

Despite the more than sixty years of focus on resonance in insect flight, previous efforts at modeling or measuring spring-wing resonance in insects have fallen short by: 1) assuming that quasi-steady assumptions on aerodynamic forces in spring-wing resonance are valid, 2) not including the effect of energetic losses from imperfect elasticity, and 3) focusing predominantly on the contributions of parallel system elasticity while disregarding contributions or limitations of series-elastic elements. Most importantly, we lack a common modeling and analysis framework to compare and contrast the energetic benefits of resonance across insects and man-made systems such as robots. Inspired by the original calculations of Weis-Fogh, we seek in this paper to develop a set of equations that govern the dynamics of parallel and series spring-wing systems using non-dimensional parameters that allow for comparative examination.

In experiments we will measure how unsteady aerodynamic effects, specifically added mass and wing-wake interaction, influence the resonant behavior of a flapping wing at hover. In order to achieve this, we compare simulations of a spring-wing oscillator subject to quasi-steady aerodynamic forces to a robophysical model of a flapping wing with known mechanical parameters subject to real fluid forces. Here we draw upon the work of van den Berg and Ellington \cite{van_den_berg_vortex_1997}, Sane and Dickinson \cite{sane_aerodynamic_2002}, and others to use a flapping wing robot that is dynamically scaled to that of flying insects. Unlike those studies, we do not prescribe the wing kinematics, instead using an elastic element in series between the wing and motor to observe resonant dynamics and emergent wing kinematics produced by varying actuation parameters. From our experiments we will develop a model to understand how structural damping influences spring-wing dynamic efficiency using non-dimensional parameters. These efforts will provide a general understanding of how springs, wings, and body mechanics converge to enable energy efficient flapping motion as a function of morphology and wing kinematics. 

\section{Theoretical preliminaries: Assumptions and Motivation}

To contextualize our study of spring-wing dynamics we first seek to outline the basic concepts of spring-wing systems. We will derive the equations of motion in the presence of aerodynamic and internal damping forces. We conclude this section with a non-dimensional representation of the equation of motion which produces two important parameters in spring-wing systems.

\subsection{Undamped parallel and series wing-spring systems}

The anatomies of flapping wing animals feature a wide range of elastic element configurations that contribute to their flapping wing dynamics. These arrangements can be expressed as combinations of springs in series and parallel with a moving inertial element, the wing (Fig.~\ref{fig:spring_wing_intro}a). In both cases, the wing interacts with a time- and history-dependent nonlinear force from the surrounding fluid. To simplify the modeling of spring-wing systems, we consider the system as a one degree of freedom rotational joint (to emulate the wing hinge), where the joint angle $\theta$ is the wing angle along the stroke plane. If we neglect internal damping, the equation of motion for a parallel spring-wing system about this joint is
\begin{equation}
    I_t\ddot{\theta}(t) + k_{p}\theta(t) + Q_{aero}(t) = T_m(t)
    \label{eqn:parallel_nodamp}
\end{equation}
where $I_t$ is the total system inertia, $k_{p}$ is the stiffness of the parallel elastic element, $Q_{aero}$ is the aerodynamic drag torque, and $T_m(t)$ is the time dependent torque applied to the wing. In the parallel configuration, the spring undergoes the same displacement as the wing and the muscle (actuator) acts directly on both the mass and spring. Nearly all spring-wing modeling of the insect thorax \cite{weis-fogh_quick_1973} and micro-aerial vehicles \cite{Whitney2012-zt, Jafferis2016-bw, Zhang2017-we,Campolo2014-tk, Baek2009-lm} have considered the parallel spring arrangement where muscle (actuator) is in parallel arrangement with the spring.
 
In a series-elastic spring-wing system, a spring is placed between the actuator and the wing. Series elastic elements are well studied in vertebrate biomechanics as muscle-tendon units where a tendon is placed between the muscle and output \cite{Roberts2011-vb}. Series elastic elements in flapping wing flight may similarly be found in the muscle tendon units of birds \cite{Tobalske2008-ie, ingersoll_how_2018}. In insects, series elasticity can arise from elastic tendons \cite{Baumler2019-xs}, elasticity in the wing hinge \cite{weis-fogh_rubber-like_1960} or within the flight power muscle \cite{Josephson2000-gt}. For simplicity of experiment design and to examine the differences between series and parallel systems, we analyze the series elastic spring-wing configuration. The equation of motion for a simple series elastic system may be written:

\begin{align}
    I_t\ddot{\theta}(t) + Q_{aero}(t) &= k_{s}(\phi(t) - \theta(t))
    \label{eqn:series_nodamp}
\end{align}

\noindent
where the force acting on the wing arises from the displacement of input angle $\phi(t)$ relative to the wing angle $\theta$. The difference between the angles is the deflection of the series spring with stiffness $k_{s}$. When the system is at steady-state (hovering), the series and parallel cases can be treated equivalently by rearranging equation~\ref{eqn:series_nodamp} to reflect the parallel configuration:

\begin{align}
    I_t\ddot{\theta}(t) + k_{s}\theta(t) + Q_{aero}(t) &= k_{s}\phi(t) 
    \label{eqn:series_as_parallel}
\end{align}

\subsection{Aerodynamic drag torque and added mass inertia}
The wing experiences an aerodynamic resisting torque, $Q_{aero}$, that opposes wing movement through the fluid. To make analysis of this system tractable, we will use a quasi-steady blade element estimate of aerodynamic torque consistent with previous quasi-steady methods for spring-wing systems \cite{weis-fogh_quick_1973} and micro-aerial vehicles \cite{Whitney2012-zt, Jafferis2016-bw, Zhang2017-we,Campolo2014-tk, Baek2009-lm}. Following the standard conventions for the quasi-steady blade element method we express the aerodynamic drag torque:

\begin{align}
Q_{aero} &= \underbrace{\frac{1}{2} \rho C_{D}(\alpha) \frac{R^5}{\AR} \hat{r}^{3}_{3}}_{\Gamma} |\dot{\theta}|\dot{\theta} \nonumber \\
 &= \Gamma |\dot{\theta}|\dot{\theta}
 \label{eqn:Torque}
\end{align}

The variable $\Gamma$ is the drag torque coefficient and is a function of wing geometry (span, $R$; aspect ratio, $\AR$; non-dimensional radius of the third moment of wing area, $\hat{r}_{3}$); pitch angle, $\alpha$; drag coefficient, $C(\alpha$); and fluid density, $\rho$. Since we are considering aerodynamic torque about the wing hinge, $Q_{aero}$ has units of [N~m] and $\Gamma$ has units of [N~m~s$^2$].

In addition to aerodynamic torque, the acceleration of a wing within a fluid leads to an additional inertia; an ``added'' or ``virtual'' mass, as it is sometimes called \cite{ellington_aerodynamics_1984-4}. We use a method from \cite{ellington_aerodynamics_1984-4} for modeling the mean added mass inertia, $I_A$ (See supplementary \S \ref{suppmat:addedmass}). The total inertia is computed as $I_t = I_{sys} + I_A$, where $I_{sys}$ is the inertia of the wing and wing transmission. In the insect flight system, the wing hinge acts as a mechanical transmission, converting linear muscle actuation to angular wing motion. It is possible that additional inertial terms from the reflected inertia of the oscillating muscle and thorax may be important. However, there is little known about the specific motion and inertia of the wing hinge transmission and the role of muscle and thoracic inertia, so we will disregard their effects in this manuscript.

\subsection{Structural damping in the insect thorax}
Recent experiments to measure the damping response of the hawkmoth (\textit{Manduca sexta}) thorax \cite{gau_indirect_2019} and cockroach (\textit{Blaberus discoidalis})  leg joints \cite{Dudek2006-rz} have both identified structural a.k.a. hysteretic damping as the dominant source of energy loss. Consistent with these observations, we seek to consider the effects of structural damping on spring-wing system dynamics. Structural damping is a common source of energy loss in biomaterials \cite{Gralka2015-vl} that differs from viscous damping in that there is no velocity dependence in the structural damping force. For general oscillatory motion, the structural damping force can be included as a modification to the spring constant, $K = k(1 + i \gamma)$, where $k$ represents either the parallel or series spring. The coefficient $\gamma$ is the structural damping loss factor \cite{beards_engineering_1995} which has been found to be $\gamma = 0.2$ for cockroach leg joints \cite{Dudek2006-rz} and $\gamma = 0.1$ for the hawkmoth thorax \cite{gau_indirect_2019}. For constant sinusoidal motion at a single frequency, $\omega$, the structural damping force can be represented as a viscous-like force with a coefficient that scales with frequency:

\begin{align}
Q_{struct} = \frac{\gamma k}{\omega} \dot{\beta}
\label{eqn:struct}
\end{align}

\noindent where the angular velocity $\dot{\beta}$ is the relative speed of spring compression (for parallel system $\dot{\beta} = \dot{\theta}$ and series systems $\dot{\beta} = (\dot{\phi} - \dot{\theta}$).
The presence of $\omega$ in the denominator makes the structural damping force frequency-independent, unlike typical viscous damping (See supplementary \S~\ref{suppmat:StructuralDamping} for derivation).

\subsection{An organizational framework for spring-wing systems}

In this final section of the motivation we introduce a set of non-dimensional variables that govern general spring-wing dynamics. As discussed in the introduction, the Weis-Fogh  number $N$ is the ratio of maximum inertial force compared to maximum aerodynamic force over a cycle. Assuming sinusoidal wing motion with amplitude $\theta_0$ and frequency $\omega$, the maximum inertial torque is $I_{t} \theta_0 \omega^2$ and maximum aerodynamic torque is $\Gamma \theta_0^2 \omega^2$ resulting in the Weis-Fogh number
\begin{align}
N &= \frac{I_{t}}{\Gamma \theta_0}
\label{eqn:WF}
\end{align}
This quantity should dictate how important a role spring elements can play in energetic efficiency at hover. For $N < 1$, aerodynamic forces dominate and kinetic energy may be fully dissipated into the surrounding fluid over each wing stroke; no elastic storage is needed. However, for $N>1$, excess kinetic energy from the wing can be recovered by a spring. Observations from biological and robotic flapping wing flyers indicate that N roughly varies between 1 and 10 for a broad range of insects \cite{weis-fogh_quick_1973}.

In order to compare across insect species, we non-dimensionlize the dynamics, assuming the wing oscillates sinusoidally at a frequency $\omega$, and with amplitude $\theta_0$. We begin with the full dynamics equation for the parallel spring-wing system including structural damping:

\begin{equation}
    I_t\ddot{\theta} + k_p\theta + \Gamma |\dot{\theta}|\dot{\theta} + \frac{\gamma k}{\omega}\dot{\theta} = T(t)
    \label{eqn:full_parallel}
\end{equation} 

\begin{figure*}[t]
    \begin{centering}
    \includegraphics[width=.9\linewidth]{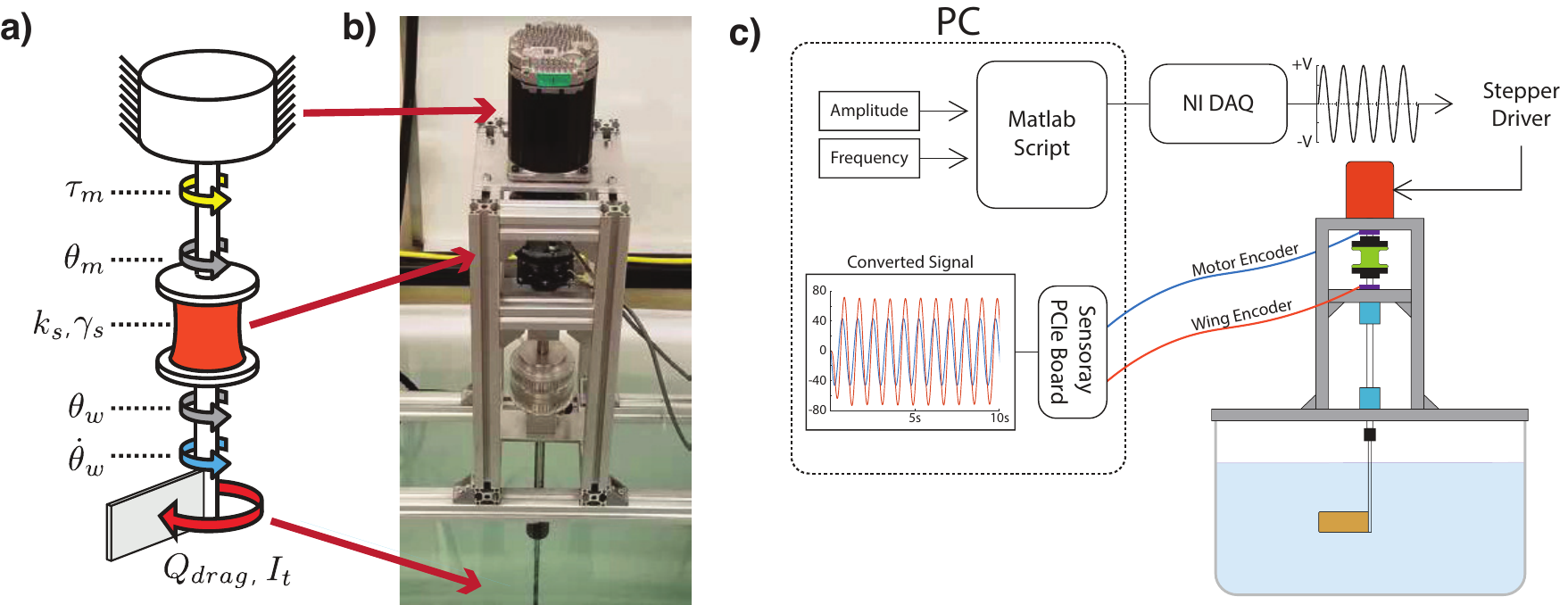}
    \caption{Schematic and details of the robophysical flapping wing experiments. a) A schematic of the system showing the stepper motor in series with a torsional spring connected to a rigid wing. b) A picture of the physical system. c) A diagram of the  control and experimental operation of the robophysical system.}
    \label{fig:robophysical}
    \end{centering}
\end{figure*}

\noindent We introduce dimensionless wing angle $q$ and dimensionless time $\tau$ and substitute into Eq. \ref{eqn:full_parallel} (See supplementary \S~\ref{suppmat:nondim}). We arrive at the following dimensionless equation of motion for the spring-wing system

\begin{flalign}
\ddot{q}_w + \hat{K} q_w + \gamma \hat{K} \dot{q}_w + \frac{1}{N} |\dot{q}_{w}|\dot{q}_{w} = \hat{T}_p(t)
\label{eqn:parallel_nondim}
\end{flalign}

\noindent
where we have defined the reduced parallel stiffness,

\begin{equation}
\hat{K}_{p} = \frac{k_p}{I_t \omega^2}
\label{eqn:k_hat}
\end{equation}

\noindent
and the Weis-Fogh number, $N$ is in the coefficient of the aerodynamic torque. The normalized torque in the parallel system is 
\begin{equation}
\hat{T}_p(t) = \frac{T(t)}{I_t \theta_0 \omega^2}
\label{eqn:normTorqueParallel}
\end{equation}

Performing a similar substitution for the series system we arrive at the equation below
\begin{align}
    \ddot{q}_w + \hat{K}_s q_w + \gamma \hat{K}_s \dot{q}_w + \frac{1}{N} |\dot{q}_{w}|\dot{q}_{w} = \hat{T}_s(t)
    \label{eqn:series_nondim}
\end{align}
\noindent
where the normalized torque in the series system is

\begin{equation}
\hat{T}_s(t) = \frac{\hat{K}_s}{\theta_0}\left(\phi(t) + \frac{\gamma}{\omega}\dot{\phi}\right)   
\label{eqn:normTorqueSeries}
\end{equation}
\noindent
We provide a full derivation of these equation in the supplementary material
(\S~\ref{suppmat:nondim}).

Through the change of variables in the parallel and series spring arrangements we have arrived at nearly two identical non-dimensional dynamics equations in Equations~\ref{eqn:parallel_nondim} \& \ref{eqn:series_nondim}. The differences between the series and parallel systems in this form are all contained in the actuation variables, $\hat{T}_s$ and $\hat{T}_p$. Thus, while the forced actuation of these system may result in different dynamics, the similar structure of the non-dimensional dynamics equations indicates that in both systems three variables likely govern the behavior: $N$, $\hat{K}$, and $\gamma$. As an exemplary demonstration of this, if either the series or parallel system is driven to steady-state oscillation and then the external actuation is turned off ($T = 0$ or $\phi = 0$), both equations~\ref{eqn:parallel_nondim} \& \ref{eqn:series_nondim} become equivalent. It is the objective of this paper seek to understand how $N$, $\hat{K}$, and $\gamma$ influence dynamic efficiency and resonance of a simple series spring-wing system.

\section{Experimental and numerical methods}
\subsection{Robophysical experiment design}
We developed an experimental spring-wing system to study flapping wing behavior in the context of realistic fluid forces (Fig.~\ref{fig:robophysical}). The quasi-steady modeling presented in the previous section greatly simplifies the unsteady, history-dependent aerodynamic phenomena involved in flapping flight. A dynamically-scaled physical wing serves as a reference against which we can evaluate the performance of the quasi-steady model. The system consists of a servo motor capable of accurate position control, a molded silicone torsional spring with linear elasticity and structural damping, and a simple fixed-pitch wing element attached to a rotary shaft and submerged in water.

\subsubsection{Motor selection and system friction reduction}

A high-torque servo motor (Teknic Clearpath SDSK) was chosen to drive the system under closed loop angular position control. The servo is able to provide substantially more torque than that experienced by the wing in the fluid, effectively decoupling the motor and wing dynamics. We monitor the motor and wing angle using two optical encoders (US Digital, 4096 CPR). To reduce the influence of friction on the wing motion we used two radial air bearings (New Way, \#S301201) which resulted in negligible bearing friction. The shaft was supported vertically by an axial thrust bearing, which did contribute a small amount of friction.

\subsubsection{Reynolds number scaling}
To ensure that we match the aerodynamic behavior of small insect wings we chose experimental parameters to dynamically scale our system. Consistent with previous dynamically-scaled experiments \cite{dickinson_wing_1999, sane_aerodynamic_2002}, we sought to maintain a Reynolds number in the range of  that of small flapping wing insects, $Re = 100-10,000$. We define Reynolds number based on standard methods, using wingtip velocity as the flow speed and wing chord as the characteristic length \cite{dickinson_wing_1999}. We choose water as a working fluid ($\rho = 997 kg/m^3$) and chose wing geometry (rectangular, 10 x 3.6 x 0.5 cm) and a range of actuation parameters (10-64 deg amplitude, 0.5 to 4.1 Hz frequency) where the resulting wing kinematics had $Re \approx 10^3 - 10^4$. Note that since the wing amplitude is an emergent property of the system due to the series spring configuration, so too is the Reynolds number of an individual test.

\begin{figure*}[t]
\begin{centering}
\includegraphics[width=1\linewidth]{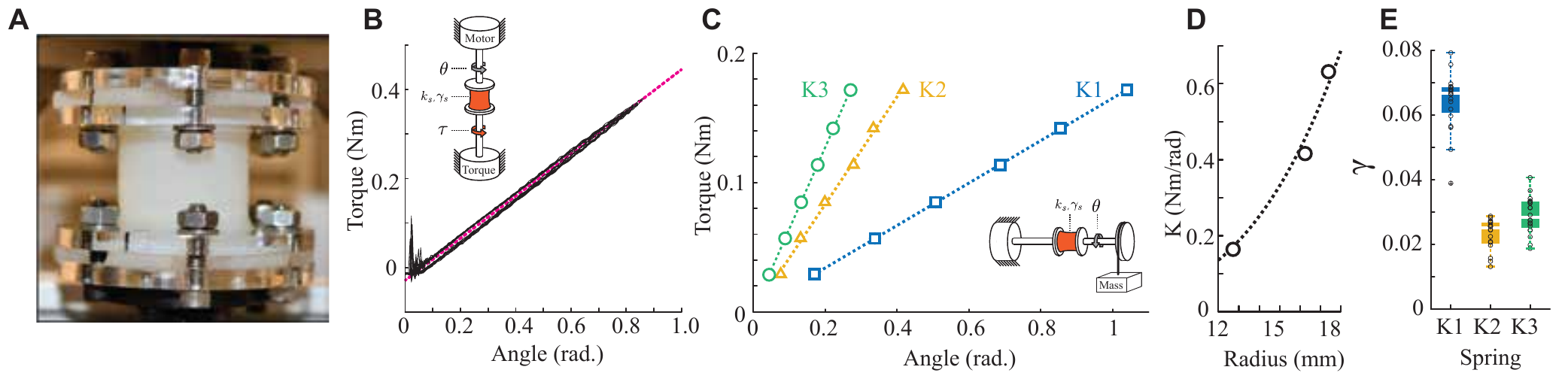}
\caption{Characterization of silicone torsion springs. (A) Photo of silicone torsion spring. Acrylic shaft couplers at the top and bottom clamp to the flanged ends of the spring. (B) Dynamic loading of the torsion spring (blue) results in a linear torque response with slight hysteresis effects. (C) Static load test results for three silicone springs. (D) Spring stiffness increases with the radius of the cylindrical region and follows an $R^4$ curve. (E) Estimates of the hysteric loss factor for 3 springs.}
\label{Fig:spring_mechanics}
\end{centering}
\end{figure*}

\subsubsection{Weis-Fogh number scaling}
In order to achieve a wide range of Weis-Fogh ~number (Eqn.~\ref{eqn:WF}) in experiment, we must be able to adjust the ratio of peak inertial torque to peak aerodynamic torque. We chose to keep the drag torque coefficient $\Gamma$ constant, so we vary $N$ by changing wing inertia and wing amplitude $\theta_0$. Wing inertia, $I_t$ was varied by adding acrylic and aluminum discs to the wing shaft, leading to total inertias of $I_t$ = [7.0, 11.5, 23.0]$\times10^{-3}$ kg~m$^2$. Our experiments resulted in a range of $N$ between 1 and 5 when operating at the resonant frequency, consistent with many insects \cite{weis-fogh_quick_1973}.

\subsubsection{Aerodynamic calculations}
We used a rigid rectangular wing with a fixed vertical pitch ($\alpha=0$), which simplifies modeling and motor control, and eliminates any energy storage and return from a flexible wing. The drag torque coefficient for a rectangular wing from Equation~\ref{eqn:Torque} is
$\Gamma = \text{1.07}\times\text{10}^{-3}$~kg~m$^2$, where the coefficient of drag is constant at $C_D(0) = $ 3.4 (taking this value from \cite{sane_aerodynamic_2002}). We also compute added mass inertia, $I_A$, for a calculated value of $3.465\times10^{-4}$~kg~m$^2$ (See supplementary \S \ref{suppmat:aerodynamic} for derivation).

\subsubsection{Silicone spring fabrication and evaluation}

We used custom fabricated silicone torsion springs for the series spring element (Fig.~\ref{Fig:spring_mechanics}a). Silicone was chosen because it can be cast into custom shapes and has a linear elastic response over large strain. The springs were designed with a cylindrical profile with flanges on each end to facilitate coupling to the motor and wing shafts (Fig.~\ref{Fig:spring_mechanics}a). Detailed information about the design and fabrication process may be found in \S \ref{suppmat:springs}.

\begin{figure*}[t]
\begin{centering}
\includegraphics[width=\linewidth]{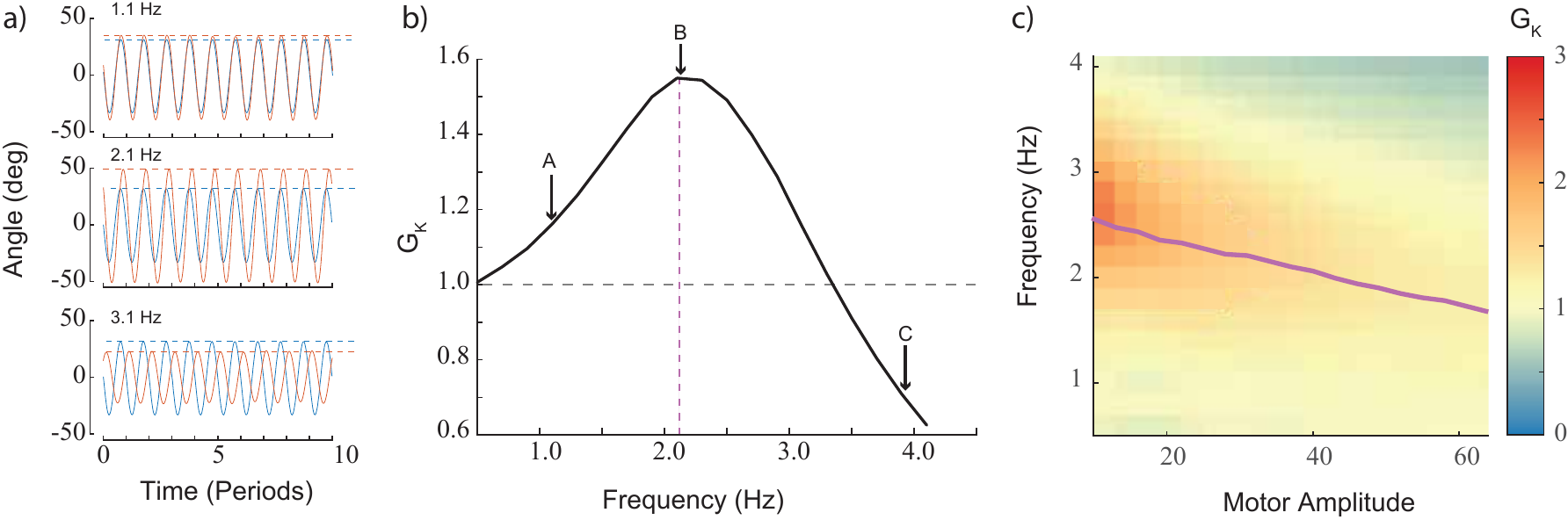}
\caption{Overview of data processing method. a) Raw data in the form of encoder readings of the motor (blue) and wing (orange) positions are collected for each combination of actuation frequency and amplitude. The three graphs show that for the same motor actuation amplitude, the wing amplitude changes with frequency. b) We define kinematic gain ($G_K$) as the ratio between wing and motor amplitude, plot change in $G_K$ over frequency, and identify peak $G_K$ and resonance frequency. The plot in (b) is the resonance curve for motor amplitude = 31 deg, [$K$,$I_t$] = [0.416 Nm rad$^{-1}$, 11.5$\times 10^{-3}$ ~kg m$^2$]. c) Plot of  $G_K$ across all motor amplitudes and frequencies with the same mechanical parameters. Peak $G_K$ and resonance frequencies fall along the purple line.}
\end{centering}
\end{figure*}

We used three spring designs with torsional stiffness values of $K_s$ = [0.163, 0.416,0.632]~Nm/rad. Figure~\ref{Fig:spring_mechanics} shows the results of experiments to characterize the spring mechanical properties. We subjected springs to both cyclic and static loading conditions (Fig.~\ref{Fig:spring_mechanics}a-b) to measure their elastic and damping properties. The spring torque response was linear over the range of angles tested (Fig.~\ref{Fig:spring_mechanics}b-c) with stiffness values that are consistent with the predicted torsional stiffness (Fig.~\ref{Fig:spring_mechanics}d). In dynamic testing we observed a small amount of hysteresis in cyclic loading experiments indicating the presence of damping within the spring (Fig.~\ref{Fig:spring_mechanics}b).
We measured the structural damping coefficient, $\gamma$, via cyclic oscillation experiments. We oscillated each spring sinusoidally across frequencies between 0.5 Hz and 10 Hz and amplitudes of 10, 20, and 30 degrees. We measured the angle-torque relationship during these tests and fit a structural damping model to this data for each spring. We find that all three springs exhibited structural damping with loss moduli shown in (Fig.~\ref{Fig:spring_mechanics}e). For comparison, the loss modulus is $\gamma = 0.1$ in Hawkmoths \cite{gau_indirect_2019} and $\gamma = 0.2$ in the legs of Cockroaches \cite{Dudek2006-rz}.

\subsubsection{Actuation and Data Acquisition}
Each experiment consisted of driving the motor angle with a specified amplitude and frequency, $\phi = \phi_0 \sin{2\pi ft}$. We varied input amplitude across $\phi_0 = $[10 - 65]~deg, in increments of 5$^\circ$ and frequency across $f$ = [0.5 - 4] Hz in increments of 0.2 Hz. Digital step and direction signals were used to set the angular position trajectory of the Clearpath servo. Each test was run for 30 seconds, during which encoder position readings were recorded from both the motor and wing by a NI PCIe board (\#6323 Multifunction IO Device), sampled at 1kHz. Encoder readings were saved as text files in Matlab and processed.

\subsection{Simulation}
We developed numerical simulations that modeled the  parallel and series configurations of the spring-wing system based off of Equations~\ref{eqn:parallel_nodamp}\&\ref{eqn:series_nodamp}. The parameters of the simulations were based on measured and calculated parameters from the experiment. Simulations were performed in Matlab using the ODE45 numerical integration function. The integration algorithm performs an adaptive step integration with absolute and relative error tolerances set at $10^{-3}$ and $10^{-6}$, respectively. 

\subsection{Data Analysis}
We performed identical analysis of both simulation and experimental data. We collected wing and motor angle data, computed wing and motor angular velocity, and found the amplitude, frequency, and relative phase of each by fitting the data to functions $\theta(t)$ and $\phi(t)$, respectively:

\begin{align}
    \phi(t) &= \phi_o \sin(\omega t)\\
    \theta(t) &= \theta_o \sin(\omega t + \psi)
\end{align}

\noindent 
See \S \ref{suppmat:data_analysis} for details. 

In order to identify the resonant peak, we define a non-dimensional term, \textit{kinematic gain}, which is the ratio of motor angle amplitude and wing angle amplitude:
\begin{equation}
    G_K = \frac{\theta_o}{\phi_o}
\end{equation}
For a resonant system, kinematic gain is maximum at the resonant frequency, $\omega_r$. We also compute the quality factor of the oscillation behavior, $Q$, using the following definition
\begin{equation}
    Q = \frac{\omega_r}{\Delta \omega}
    \label{eqn:quality}
\end{equation}
\noindent
Where $\omega_r$ is the resonant frequency and $\Delta \omega$ is full width at half maximum. The quality factor is a metric of the sharpness of the resonant peak: high $Q$ means kinematic gain, $G_K$ drops off as $\omega$ moves away from the resonant frequency, while low $Q$ means $G_K$ changes slowly with varying $\omega$.

In order to identify $\omega_r$ for a set input amplitude experiment, we locate the frequency that maximizes kinematic gain. We fit a 5th order polynomial to the 12 points closest to the peak measured $G_K$ to get a smooth approximation of the resonance curve. The maximum value of the polynomial is the peak gain, and the frequency corresponding to it is the resonant frequency. When reporting $N$, dynamic efficiency, etc. ``on resonance," we use the experimental configuration with a frequency closest to the resonant frequency. As a result, some nominally resonant points are not exactly on the resonant peak, but may be off by as much as 0.25 Hz.

We use the position measurements, their derivatives, and the known mechanical parameters of the system to estimate the torques on the system: Aerodynamic, inertial, elastic, motor, and structural damping.
\begin{align}
    T_{aero}(t) &= \Gamma |\dot{\theta}(t)|\dot{\theta}(t)\\
    T_{inert}(t) &= I\ddot{\theta}(t)\\
    T_{elast}(t) &= -T_{motor} = k_s(\phi(t)-\theta(t)) \\
    T_{d}(t) &= \frac{k_s\gamma}{\omega}\left(\dot{\phi(t)} - \dot{\theta}(t)\right)
\end{align}
Note that we compute the equivalent non-dimensional torques and kinematics using the terms defined in Equations~\ref{eqn:parallel_nondim}~\&~\ref{eqn:series_nondim}.

Lastly, we compute the dynamic efficiency of the oscillatory motion. The dynamic efficiency is defined as the ratio of aerodynamic work to input work
\begin{align}
    \eta &= \frac{W_{aero}}{W_{tot}}
\end{align}
over a stroke (or equivalently a half-stroke). To calculate the aerodynamic and total work in experiment we use the following equations
\begin{align}
    W_{aero} &= \int_{-\theta_o}^{\theta_o} T_{aero} \; d\theta \\
    W_{tot} &= \int_{-\theta_o}^{\theta_o} R(T_{motor}) \; d\theta
\end{align}
Note that in the motor work, $R(x)$, is a rectification function defined as $R(x) = x$ for $x > 0$ and $R(x) = 0$ for $x < 0$. This rectification function accounts for the fact that negative mechanical work in insect flight muscle does not contribute significantly to metabolic energy expenditure \cite{weis-fogh_quick_1973,Dickinson1995-ot}.

\section{Results}

\subsection{Kinematic gain varies with actuation and system properties}

\begin{figure}
    \centering
    \includegraphics[width=1\columnwidth]{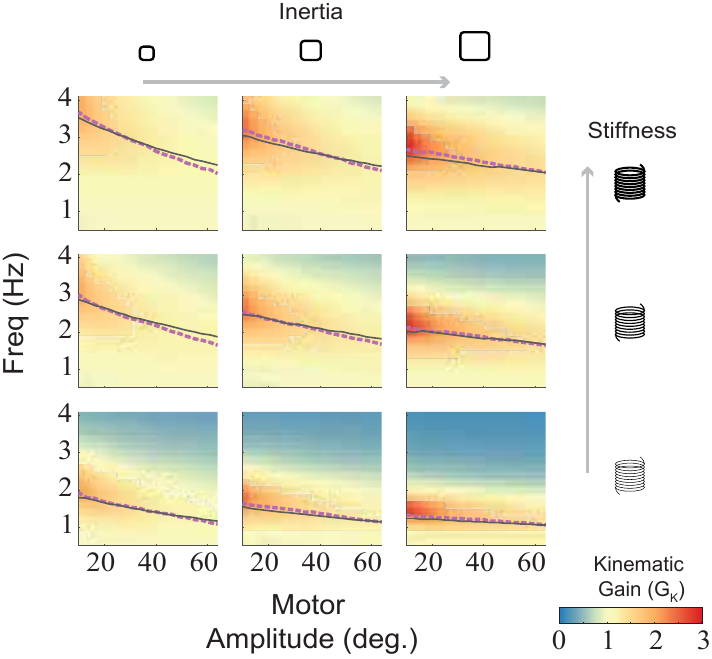}
    \caption{Results across all springs, inertias, actuation parameters. Peak $G_K$ are plotted in purple dashed-lines. Peak $G_K$ from simulation using measured system parameters and identical actuation are shown in solid black lines.}
    \label{Fig:expResultsPlots}
  \end{figure}

We performed 3249 tests with varying input parameters (amplitude and frequency) and mechanical system parameters (spring stiffness and inertia) and measured the emergent wing kinematics (Fig \ref{Fig:expResultsPlots}). Results for the nine inertia and stiffness combinations are shown as heatmaps with color indicating kinematic gain. The arrows on the right and top of the figure denote the directions of increasing stiffness and inertia, respectively.

For all stiffness and inertia combinations, we observed that the peak $G_K$ occurred at the lowest actuation amplitudes. At low amplitudes, aerodynamic damping is minimized and less energy is lost to the surrounding fluid, allowing for higher kinematic gain. The maximum kinematic gain increased with increasing system inertia in all cases, reaching a maximum when the system is a combination of a soft spring and high inertia (bottom right corner).

The resonant frequency decreased as the motor amplitude (and thus the emergent wing amplitude) increased. The dashed purple line in Figure~\ref{Fig:expResultsPlots} shows peak kinematic gain for each motor amplitude. This decrease in resonant frequency is consistent with simulation predictions shown as solid black lines in Figure~\ref{Fig:expResultsPlots}. We discuss this model in the first discussion section below.

\subsection{Flapping resonance with quasi-steady aerodynamics}

Following the experiments, we sought to see how much of the observed dynamics is predicted by the simplified, quasi steady equations of motion described in Section 2. The real aerodynamic loads on the wing are time- and history-dependent, so it is not clear to what degree those unsteady loads affect the resonance properties of the system at steady state.

\begin{figure*}
    \centering
    \includegraphics[width=1\textwidth]{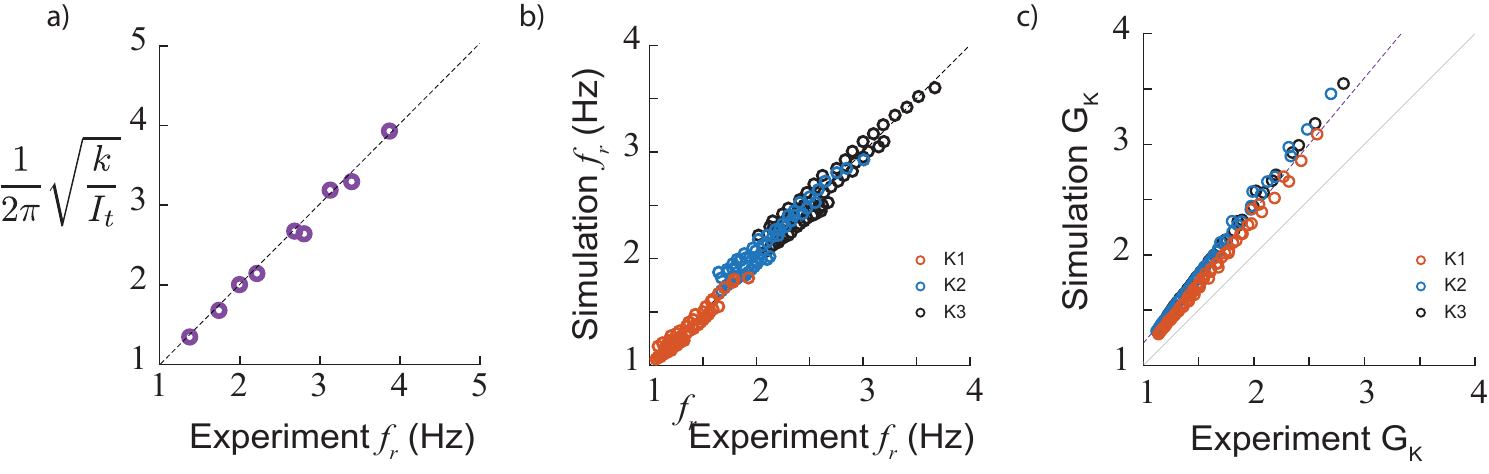}
    \caption{a) Comparison of system natural frequency from calculations to measured low-amplitude resonant frequency in oscillation experiments. b) Comparison of simulated resonance frequency to experimental. c) Comparison of simulated kinematic gain to experimental}
    \label{fig:comparison}
  \end{figure*}
\subsubsection{Natural frequency of the system matches lightly-damped resonant frequency}
At low motor amplitude, the system dynamics are only weakly affected by the aerodynamic force and thus the resonant frequency should be determined by the natural frequency of the spring-wing system. Comparison of the experimentally measured resonance frequency at the lowest amplitude with natural frequency computed from spring stiffness and inertia displays extremely good agreement (Fig.~\ref{fig:comparison}a). Thus our intuition is confirmed that low amplitude wing oscillation leads to small aerodynamic damping and the system better resembles a weakly-damped spring-mass oscillator.

\subsubsection{Quasi-steady simulation predicts experimental resonant frequency}
As motor and wing amplitude increase, we observe reductions in both kinematic gain and resonant frequency. This behavior is consistent with the quadratic velocity damping generated by the aerodynamic force on the wing and has been observed in other spring-wing experiments \cite{Jafferis2016-bw}. We sought to determine if a quasi-steady aerodynamics model was sufficient to model the spring-wing behavior observed in experiment. By performing numerical simulations with identical parameters as the experiments of Fig.~\ref{Fig:expResultsPlots} we measured the simulated resonant frequency across all amplitudes. We find that the quasi-steady simulation resonant frequency agrees well with the experimental resonance frequency (Fig.~\ref{fig:comparison}b). This suggests that quasi-steady aerodynamic modeling is sufficient to capture the spring-wing relationship between amplitude and resonance frequency across all motor amplitudes.

\subsubsection{Quasi-steady simulation over-predicts kinematic gain}
Despite the agreement in resonant frequency, we observed a difference between the kinematic gain in the experiment and the simulation (Fig.~\ref{fig:comparison}c). The simulation over-predicted the kinematic gain for all resonance experiments by a maximum of 20\% at the highest experimental gains. The over-prediction error grew with increasing gain, suggesting a systematic error in the modeling of the spring-wing system. The combination of disagreement between simulation and experiment in kinematic gain and the good agreement in resonant frequency suggests to us that unmodeled dissipation from system friction is likely the cause. Coulomb friction in the bearings would not influence the system resonant frequency, but would decrease the output amplitude, consistent with our observations. We opted not to include friction in our modeling for two reasons: 1) we kept only the biologically-relevant damping terms in the system equations, and 2) modeling friction can be complicated due in part to highly-nonlinear stick-slip phenomena \cite{Rice2001-wv}.

\subsection{Dynamic efficiency is amplitude and frequency dependent}

\begin{figure}
    \centering
    \includegraphics[width=1\columnwidth]{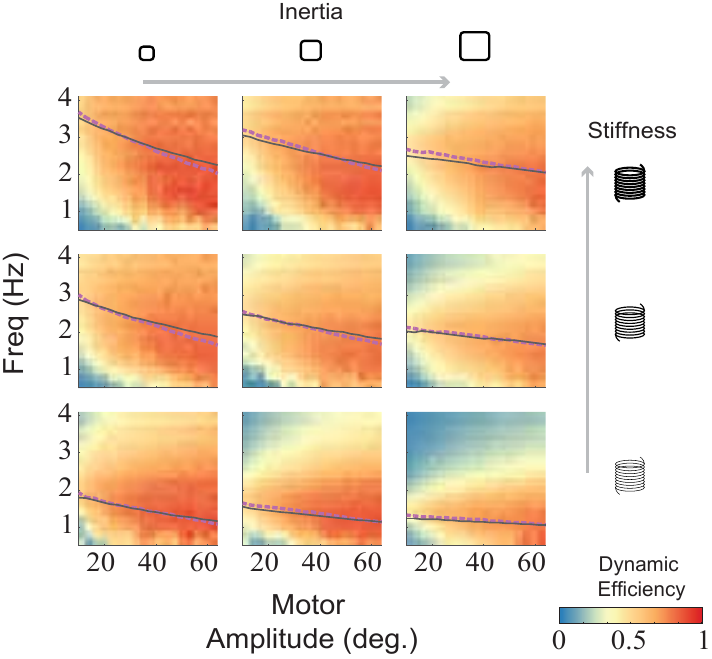}
    \caption{Dynamic efficiency as a function of motor amplitude and frequency. Experiments are ordered in increasing spring stiffness (rows) and inertia (columns) as indicated by the arrows and text. Dashed lines are experimentally determined resonance while solid lines are model predictions for resonance.}
    \label{fig:dyn_eff_experiment}
  \end{figure}

To determine the efficacy of elastic springs for energy reduction in a flapping wing system, we calculated the dynamic efficiency, $\eta$, across all system and actuation parameters (Fig.~\ref{fig:dyn_eff_experiment}). Generally we observe that dynamic efficiency is maximum along the resonance curves for all experiments. These results are consistent with the interpretation that maximum kinematic gain corresponds to maximum energetic benefits of having a spring, e.g. that the system is at resonance. Notably the dynamic efficiency is very sensitive to oscillation frequency for low motor amplitudes while higher motor amplitudes show a very broad dynamic efficiency. The results at high amplitude are consistent with the broad dynamic efficiency versus frequency curves measured in experiments on a flapping wing robot in \cite{Campolo2014-tk}.

\section{Discussion}

In this discussion, we recall the preliminary framework established in Section 2. Through a change of variables we were able to express the equations of motion of the series and parallel spring-wing systems (Equations \ref{eqn:parallel_nodamp}~\&~ \ref{eqn:series_nodamp}) with nearly identical expressions. In this discussion we now seek to interpret the series elastic experiment and simulation results above in the context of the non-dimensional variables: the reduced stiffness $\hat{K}$, the Weis-Fogh number $N$, and the structural damping coefficient $\gamma$.

\subsection{The Weis-Fogh number, $N$ governs resonant properties in spring-wing systems}

The frequency response and kinematic gain of the spring-wing system indicates that the resonant behavior of the system is dependent on the flapping amplitude and frequency. In a standard spring-mass system with a viscous damper, frequency dependent kinematic gain is to be expected. However, the dependence of the system resonant frequency on oscillation amplitude in the series spring-wing system (Fig.~\ref{Fig:expResultsPlots}) is different than that of the standard viscously-damped spring-mass system (unless the damping is close to a critical value). Similar to the way the spring-mass system is often reduced to non-dimensional ratios (damping ratio, $\zeta$ and reduced frequency, $\omega /\omega_n$) that govern the oscillation behavior, we here seek to show how the non-dimensional parameters of Equations~\ref{eqn:parallel_nondim}~\&~\ref{eqn:series_nondim} govern the resonance properties of the system.

\subsubsection{$N$ is a measure of aerodynamic damping in spring-wing resonance}
An important metric of a resonant system is the quality factor, $Q$, a measure of how damped the oscillator is and which, in experiment, is the sensitivity of kinematic gain to frequency change. For a linear spring-mass system, $Q$ is independent of oscillation amplitude, but  it is inversely proportional to the damping ratio, $\zeta$. In the non-dimensional spring-wing equations, the two terms that govern system damping are the Weis-Fogh number, $N$ (the prefactor of the aerodynamic term) and the structural damping loss modulus $\gamma$. We expect that for any useful spring-wing system, the energy loss due to aerodynamic damping should be much larger than the parasitic energy loss from internal structural damping. Thus, we are inspired to examine how quality factor varies with the Weis-Fogh number.

\begin{figure*}
    \centering
    \includegraphics[width=0.7\textwidth]{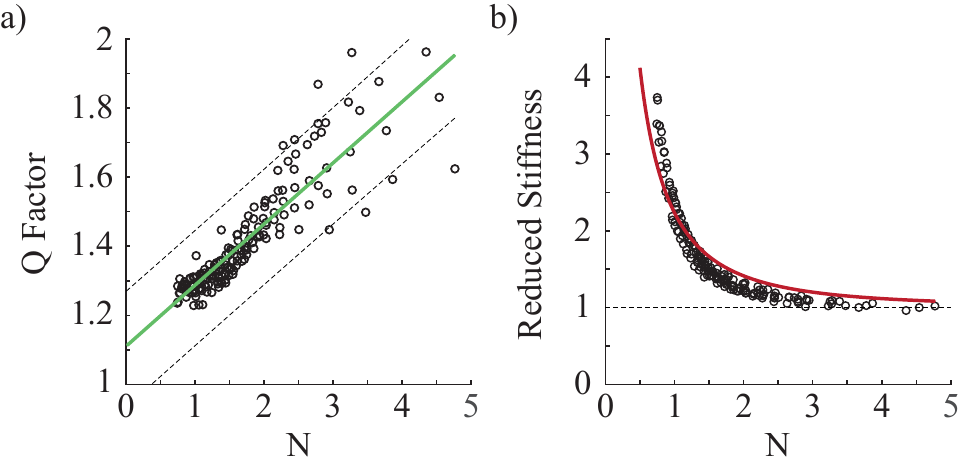}
    \caption{a) The quality factor of the oscillation as a function of Weis-Fogh number, $N$, in experiment with a linear fit and 95\% confidence interval. b) Relationship between reduced stiffness $\hat{K}$ and Weis-Fogh number, $N$ in experiment. The line represents the optimal stiffness-damping relationship (Eq. \ref{eqn:resfreq})}
    \label{fig:qfactor}
\end{figure*}

Examining the dependence of $Q$ on the Weis-Fogh number, we observe that the quality factor grows approximately linearly with the Weis-Fogh number (Fig.~\ref{fig:qfactor}a) across the ranges observed in experiment. Simulations further reveal a linear relationship between Weis-Fogh number and system quality factor given by

\begin{align} \label{eqn:QvsN}
    Q = c_1 N + c_2
\end{align}

\noindent
where $c_1 = 0.18 \pm 0.01$ and $c_2 = 1.11 \pm 0.03$ (95\% CI). By analogy with the linear spring-mass system, the linear relationship between $Q$ and $N$ reveals that the quantity $\frac{1}{N}$ is comparable to the damping ratio for linear spring-mass systems.

The importance of this relationship is that $Q$ has historically been used to evaluate the ability of insects to benefit from elastic energy storage. Ellington, using $Q$ defined as the ratio of peak kinetic energy versus energy dissipated has reported $Q$ values for a variety of insects such as the fruit-fly ($Q$ = 6.5), the hawkmoth  ($Q$ = 10), and the bumblebee  ($Q$ = 19) \cite{Ellington1999-un}. However, the experimental relationship between quality factor and $N$ we observe would suggest substantially lower quality factors. For example, from the functional relationship in equation \ref{eqn:QvsN}, we would predict quality factors for the fruit-fly (\textit{Drosophila melanogaster},  $N \le$ 1 , $Q \approx 1.2$), the hawkmoth  (\textit{Manduca secta}, $N$ = 3.6, $Q \approx 1.7$), and the bumblebee  (\textit{Bombus terrestris}, $N$ = 3.1, $Q \approx$ 1.6). Weis-Fogh numbers for the relevant animals were provided by Weis-Fogh in \cite{weis-fogh_quick_1973}. Furthermore, metabolic measurements of \textit{Drosophila m.} have indicated only 10\% energetic benefit of springs suggesting a substantially lower $Q$ than that predicted by Ellington. \cite{dickinson_muscle_1995}. This may partially be explained by the fact that the two methods of computing $Q$ are not exactly equivalent, but unfortunately, we are not able to directly compare parameters because the calculations are not reported in the previous manuscript.

\subsubsection{The resonant frequency of a series spring-wing system varies with $N$}
Examination of the Weis-Fogh number shows that $N$ grows inversely with wing amplitude (Equation~\ref{eqn:WF}). Thus, experiments that were performed at low motor amplitude correspond to spring-wing systems with large $N$. This observation, coupled with the insight from above, immediately provides understanding for why the resonant frequencies at low amplitudes match the system's natural frequency (Fig.~\ref{fig:comparison}a). At low amplitude (high $N$) the influence of damping is very small (scales as $\frac{1}{N}$) and thus increasing $N$ in spring-wing systems results in minimizing the the effects of aerodynamic nonlinearity and energy loss.

However, this insight does not fully explain why the resonance frequency should change with amplitude. Examination of the relationship between the non-dimensional variables that relate frequency ($\hat{K}$) and damping ($N$) at resonance shows a tight locus of points that indicate a potential relationship (Fig.~\ref{fig:qfactor}b). To understand this relationship we follow the method originally introduced by Bennett et al. for the analysis of series elasticity in whale flukes \cite{bennett_elastic_1987}. In that paper, the authors determine the instantaneous aerodynamic, inertial, and elastic power contributions in the system and then minimize the total power consumption over a wing-stroke (see supplementary material \S \ref{supp:bennettcalc}) to identify an optimal combination of parameters. The combination of parameters that minimize total power is given by the following relationship:
\begin{align}
    \hat{K} = \sqrt{1 + 4/N^{2}}
    \label{eqn:resfreq}
\end{align}
We plot this relationship in Figure~\ref{fig:qfactor}b and observe good agreement ($R^2 = 0.868$, $RMSE = 0.22$) between the predicted optimal series spring-wing relationship and observed relationship at resonance. The observed differences between experiment and theory could result from measurement error, or alternatively friction, structural damping, or aerodynamic effects that are unmodeled in the original derivation. However, solving Eqn.~\ref{eqn:resfreq} in terms of the actual resonant frequency, $\omega_r$ and system parameters (see supplementary material \S \ref{supp:bennettcalc}) shows very good agreement over the frequency, motor amplitude parameter space (Fig.~\ref{Fig:expResultsPlots}, black line). For completeness, we note that for a parallel spring-wing system, the resonance relationship is constant: $\hat{K} = 1$ for all $N$ \cite{bennett_elastic_1987}.

The overall intuition from this resonance analysis is that the Weis-Fogh number plays an important role in determining the spring-wing system's resonant behavior. Furthermore, the relationship between $N$ and $\hat{K}$ at resonance implies that these two variables likely define a non-dimensional parameter space for general spring-wing systems of arbitrary morphology, mechanical properties, and actuation. In the following sections we will expand that space by exploring how $N$, $\hat{K}$, and structural loss factor $\gamma$ influence dynamic efficiency.

\subsection{Energy loss from structural damping scales with Weis-Fogh number in real spring-wing systems}

\begin{figure*}
    \centering
    \includegraphics[width=0.9\textwidth]{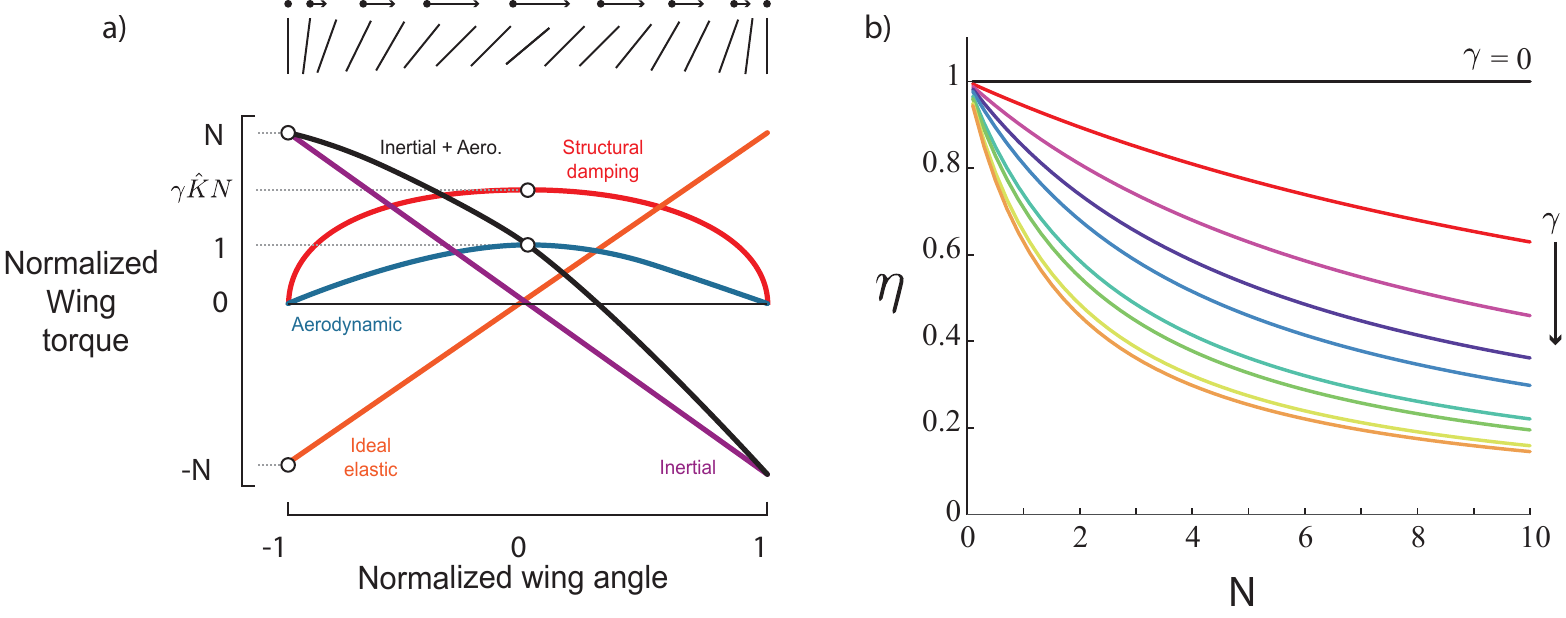}
    \caption{a) Non-dimensional torques acting on the wing hinge during the upstroke or downstroke. b) Theoretical dynamic efficiency (Eqn.~\ref{eqn:nondimwork}) for a parallel spring-wing system. Top curve is $\gamma = 0$ and curves below are increases of $\gamma$ in increments of 0.5.}
    \label{fig:structure_dyneff}
  \end{figure*}

Recent discoveries that structural damping in hawkmoth thoraxes \cite{gau_indirect_2019} and cockroach legs \cite{Dudek2006-rz} may cause significant energy loss prompts us to investigate how structural damping contributes to spring-wing energetics. 
To gain insight into energy loss scaling, we consider the parallel arrangement of a spring and a wing (Fig.~\ref{fig:spring_wing_intro}a). The parallel arrangement, as opposed to the series arrangement, is convenient because the spring (and structural damper) are subject to identical displacements as the wing hinge, and thus all forces (spring, damping, aerodynamic, and inertial) can be represented as functions of wing angle. In the series case, the actuation and wing trajectories are out of phase and therefore more complicated, if not impossible, to express analytically. In the parallel system, we can conveniently visualize all the relevant forces acting on the wing by plotting torque contributions versus wing angle (Fig.~\ref{fig:spring_wing_intro}a and Fig.~\ref{fig:structure_dyneff}a).

We now consider the scaling of each of these forces with the non-dimensional system parameters. Following the method introduced by Weis-Fogh, we normalize all the non-dimensional torques (Equations \ref{eqn:parallel_nondim} \& \ref{eqn:series_nondim}) by the peak aerodynamic torque at mid-stroke. The result is the following set of non-dimensional torques, expressed as functions of normalized wing angle, $q$. The tilde symbol represents torques normalized by peak aerodynamic torque:
\begin{align}
    &\tilde{Q}_{aero} = (1 - q^2) \nonumber \\
    &\tilde{Q}_{structural} = \gamma \hat{K} N \sqrt{1 - q^2} \\
    &\tilde{Q}_{elastic} = \hat{K} N q  \nonumber \\
    &\tilde{Q}_{inertial} = -N q  \nonumber  \nonumber
    \label{eqn:non_dim_parallel}
\end{align}
See supplementary section \S \ref{supp:parallel} for the full derivation of these terms.

In the parallel system, the ideal spring torque is exactly opposite the inertial torque and thus cancels the inertial work throughout the wing stroke, $\tilde{Q}_{elastic} = \tilde{Q}_{inertial}$. This relationship implies that $\hat{K} = 1$ in the ideal parallel system. Converting this expression back to dimensional form returns the expected relationship that defines a parallel resonant system, $k_{p} = I_t \omega^2$.

Critically for analysis of spring-wing energetics, all of these equations can be integrated over the wing stroke to provide expressions for the non-dimensional work the wing has performed. In the case of parallel resonance ($\hat{K} = 1$) we simply need to integrate the $\tilde{Q}_{aero}$ and $\tilde{Q}_{structural}$ terms since the inertial and elastic work exactly cancel. Performing these integrals over the range $q = $[-1, 1] results in the following expressions

\begin{align}
    &\tilde{W}_{aero} = \frac{4}{3} \\
    &\tilde{W}_{structural} = \frac{\pi}{2}\gamma N
\end{align}

\noindent
Since aerodynamic and structural damping work are the only sources of energy loss (and thus the only required energy input) in the system we can now express the dynamic efficiency in terms of these two energy losses.
\begin{align}
    \eta &= \frac{\tilde{W}_{aero}}{\tilde{W}_{aero} + \tilde{W}_{structural}} \\
    &= \frac{1}{1 + \frac{3\pi}{8} \gamma N}
    \label{eqn:nondimwork}
\end{align}

We now have an expression for the dynamic efficiency of a parallel spring-wing system operating at the resonance frequency. The dynamic efficiency is a function of only the Weis-Fogh number, $N$, and the structural damping $\gamma$. Examination of this expression indicates that if there is no structural damping in the system ($\gamma = 0$) the dynamic efficiency is constant, $\eta = 1$, across all $N$, indicating that all input work goes towards aerodynamic work at steady state. However, for any non-zero structural damping in the transmission, the dynamic efficiency is a monotonically decreasing function of $N$, since a portion of the input work must be diverted to overcome the internal structural damping.


\begin{figure*}
  \centering
  \includegraphics[width=\textwidth]{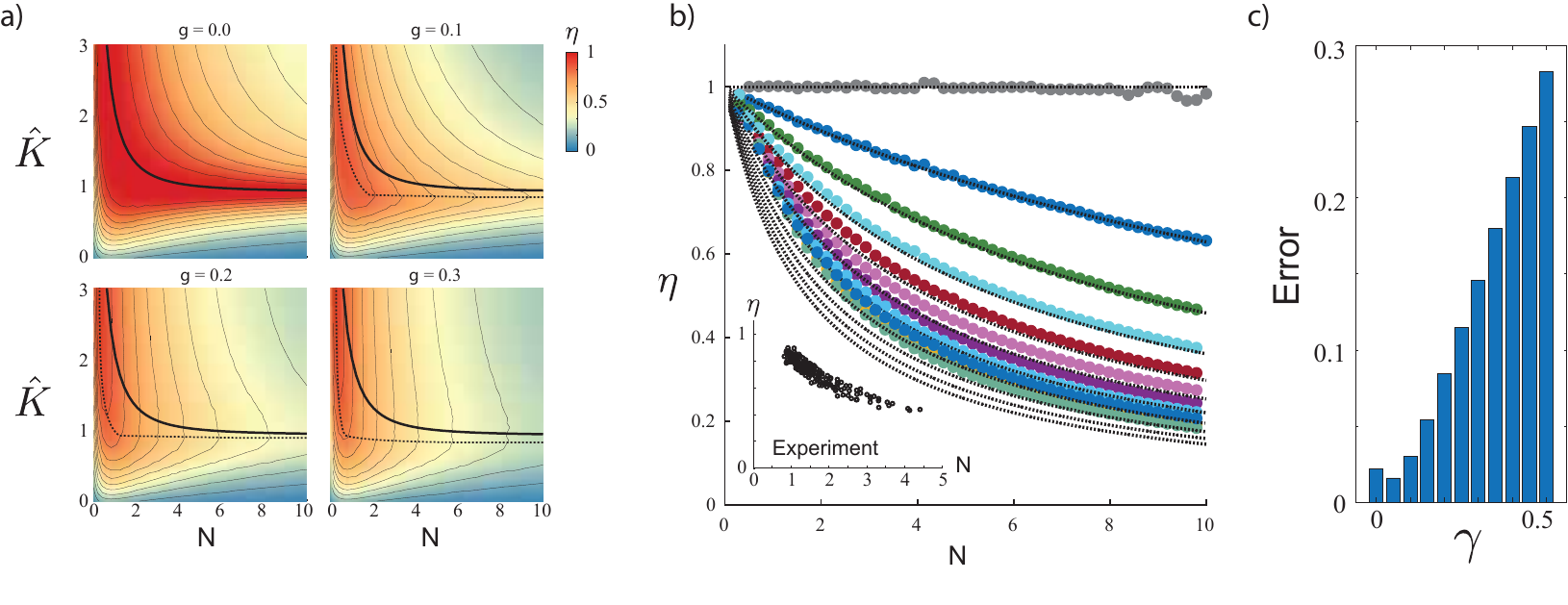}
  \caption{a) Simulation of dynamic efficiency in a series spring-wing system for four values of structural damping. The solid black line is the undamped resonance relationship between $N$ and $\hat{K}$. The dashed lines are drawn to guide the eye along the estimated minimum of the dynamic efficiency gradient. b) Numerical calculations of dynamic efficiency along the undamped resonance relationship curves for $\gamma$ = [0, 0.5] in increments of 0.05. Dashed lines show results from a parallel spring-wing system at resonance for comparison. Inset shows the dynamic efficiency of the experimental spring-wing system at resonance. c) Sum of squares difference between the series simulation $\eta$ and the parallel closed form $\eta$ for varying amounts of structural damping ($\gamma$).}
  \label{Fig:DynEff}
\end{figure*}

This analysis of a parallel spring-wing system has provided insight into how structural damping influences the mechanical efficiency of the flight transmission, i.e. dynamic efficiency. We would like to consider the same theoretical analysis for a series spring-wing system, but in this case the theoretical approach becomes intractable. In a series spring-wing system, the relative phase difference between wing and actuator changes with actuation parameters and thus the relative spring extension (and rate of extension) is not as easily determined. Bennett et. al. \cite{bennett_elastic_1987} presented an elegant reformulation of the problem to generate a closed form expression for the actuator power required in a series spring-wing system. However, this method cannot be used to include structural damping because it would require knowledge of the actuator input kinematics. Thus, to determine the influence of structural damping in a series spring-wing system we will resort to numerical methods in the next section.

\subsection{Structural damping reduces peak dynamic efficiency in series spring-wing systems}

The dynamic efficiency $\eta$ for series spring-wing systems was determined through numerical simulation of the non-dimensional equation of motion \ref{eqn:series_nondim} across a range of $N$ and $\hat{K}$. As expected in the case of no structural damping, $\gamma = 0$, the dynamic efficiency was $\eta = 1$ for all values of $N$ and $\hat{K}$ along the series resonance curve, (Eqn.~\ref{eqn:resfreq}) (Fig.~\ref{Fig:DynEff}a). These simulations also allow us to observe how $\eta$ varies across the full ($N$, $\hat{K}$) parameter space. In general, we observe stronger sensitivity of $\eta$ to changes in $\hat{K}$ as $N$ increases. Recalling our analysis of the connection between quality factor and Weis-Fogh number $N$, it is clear that, for a spring-wing system with fixed morphology and no structural damping, as $N$ increases, the variation of $\eta$ with frequency becomes more significant.

For systems with non-zero structural damping, as expected in any real system (and as observed recently \cite{gau_indirect_2019,Dudek2006-rz}), the dynamic efficiency in general decreases with increasing $N$ for all values of $\hat{K}$. For small structural damping ($\gamma = 0.1$; Fig.~\ref{Fig:DynEff}b) we observe a general similarity of the dynamic efficiency to the undamped case. The gradient of the dynamic efficiency can be observed by the spread of the contour lines, where a steep gradient is indicated by closely spaced contour lines. In the undamped case, the resonance relationship follows the line of minimum gradient in dynamic efficiency ($\text{min} \nabla \eta$): the resonance curve exactly follows the contour line of $\eta = 1$ (and thus $\nabla \eta = 0$). Examination of $\eta$ for increasing structural damping indicates the curve of minimum $\nabla \eta$ likely differs from the undamped resonance as illustrated in Fig.~\ref{Fig:DynEff}a, where the dashed lines represent estimates of the line of minimum gradient to guide the eye.

\begin{figure*}[t]
    \begin{centering}
    \includegraphics[width=.8\linewidth]{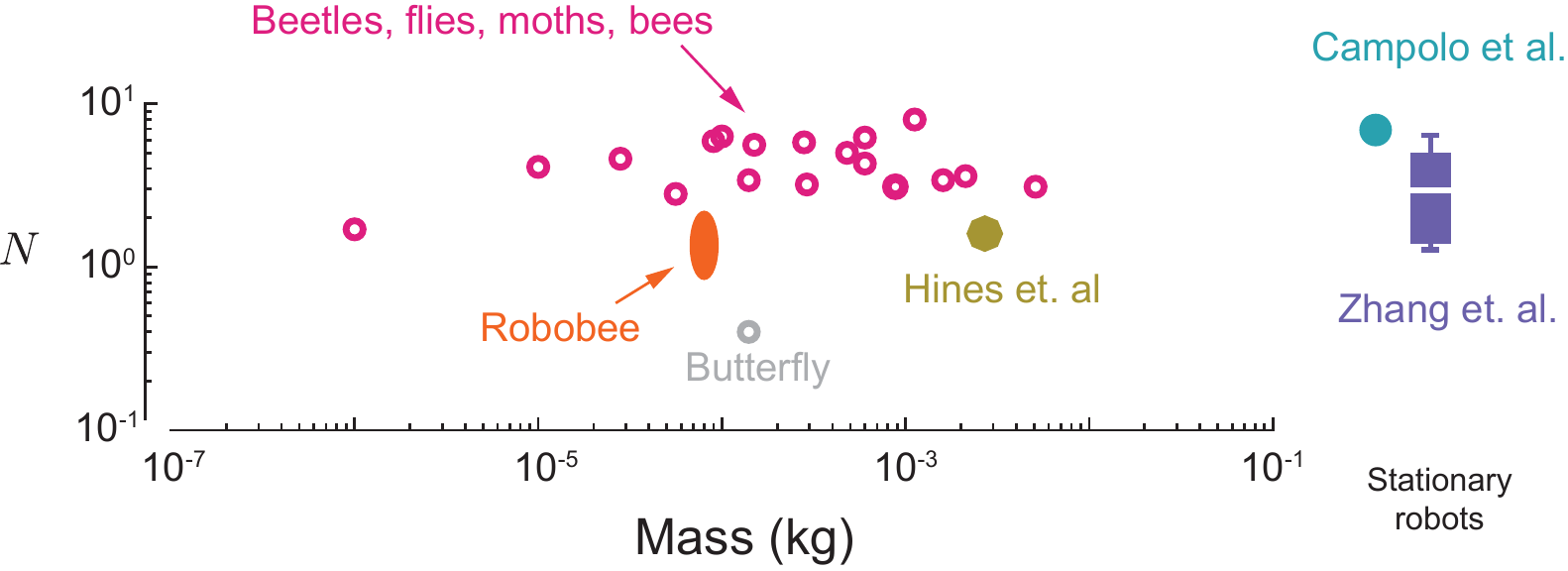}
    \caption{The Weis-Fogh number for flying insects and robots as a function of body mass (left plot) and in fixed flapping wing robots (right boxplot). Biological data is reported from \cite{weis-fogh_quick_1973}, and robot data from \cite{Jafferis2016a, Hines2014-kg, Azhar2012-hm, Campolo2014-tk}. Most observed insects and robots lie within the mass-independent range of $N \in [1, 10]$.} 
    \label{fig:N_values}
    \end{centering}
\end{figure*}
Evaluating the dynamic efficiency across the different values of $\gamma$ shows that $\eta$ monotonically decreases with $N$, consistent with our analysis of the parallel spring-wing system. In the inset of Figure~\ref{Fig:DynEff}b, we show the dynamic efficiency results at resonance for the experimental series spring-wing system. The experiment exhibits the same monotonically decreasing trend of $\eta$ with $N$. However the magnitude of $\eta$ in the experiment differs from that predicted by the simulation with structural damping alone. Thus, similar to the previous comparison of simulation and experiment (Fig.~\ref{fig:comparison}) we observe qualitatively similar trends between experiment and simulation, however the experiment exhibited lower $\eta$, likely due to additional sources of energy loss from friction and other unmodeled effects.

Overall, both the experimental and simulation results provide evidence that, for any spring-wing system with structural damping, the dynamic efficiency decreases monotonically with increasing $N$. These results are a bit counter-intuitive from the discussion above, where quality factor is observed to scale linearly with $N$ (Fig.~\ref{fig:qfactor}a). For a perfect spring-wing system ($\gamma = 0$) it is true that increasing $N$ diminishes the relative energy loss from aerodynamic work compared to the total energy of the oscillator (the definition of the quality factor). However, in the presence of internal energy losses, the actuators have to do extra work to overcome internal body damping, which scales with $N$. With internal damping, the quality factor might still be the same (since it is a ratio of energies) but the internal damping decreases the useful energy of the oscillator and thus dynamic efficiency.


\subsection{Intermediate Weis-Fogh number may balance damping losses and elastic benefits}

The scaling of dynamic efficiency with $N$ may help explain why flapping wing insects and robots tend to have an $N$ in the range of 1-10 across three orders of magnitude in body mass (Fig.~\ref{fig:N_values}). A very low $N$ means that aerodynamic forces dominate and wing kinetic energy will be dissipated by aerodynamic work rather than be stored in a spring \cite{dickinson_muscle_1995}. High $N$, however, increases the internal energy losses from structural damping and reduces the benefits of elastic energy storage. Thus, flapping wing insects and birds with Weis-Fogh number in an intermediate range may balance the positive benefits of spring-wing energy exchange with the parasitic energy losses of internal structural damping. In order to achieve high dynamic efficiency at hover, wing geometries, flapping amplitudes, and wingbeat frequencies may be tuned to maintain operation in this restricted regime of $N$s.

In addition to the constraints of damping, high $N$ biological spring-wing systems may not be possible due to biomaterial or physiological constraints. For example, high $N$ spring-wing systems that are capable of hover (implying modest wingbeat amplitude, $\theta_0$) would require large wing inertia which may be impractical due to the possible imposition of extra system weight. Similarly, from the elastic materials perspective a high-$N$ spring-wing system with large wing inertia would require extremely stiff and resilient elastic elements to operate at resonance. It may be that for extremely large $N$ spring-wings the required elastic stiffness may exceed the practical regime of biological materials. Both of these considerations however do not necessarily limit robotic systems from being developed with high $N$ through appropriate inertial and stiffness design considerations.

Lastly, it is unlikely that dynamic efficiency at hover is the only factor that dictates this range of morphologies. Other factors, such as the effect of $N$ on transient dynamics and control of wing kinematics, are likely to be significant. For example, in a spring-wing system with high-$N$ the wing dynamics are dominated by inertial effects and thus wing kinematics are likely insensitive to transient changes in aerodynamic forces. Such a high-$N$ system may have wingbeat kinematics that are relatively stable in the presence of gusts of wind. However, in the opposite case of a low-$N$ spring-wing system, the flyer would be able to more easily modulate wing kinematics and possibly wingbeat frequency for control purposes. In these cases and others, the Weis-Fogh number may provide a baseline for insight in broad comparative study of flying insects, enabling identification of commonalities between species as well as exceptional cases that merit further study.


\section{Conclusion}

Many flapping wing insects and birds possess elastic elements in their body that may reduce the power demands of flapping wing flight. However, recent experiments have demonstrated that insects are also subjected to internal power loss from the deformation of their thorax. In this manuscript we have introduced three non-dimensional variables for general spring-wing systems that govern oscillatory behavior and dynamic efficiency. Inspired by the foundational work of Weis-Fogh, we re-introduce the ratio of maximum inertial force to aerodynamic force as the Weis-Fogh number, $N$. Experiments and simulation illustrate that $N$ is a fundamental parameter of spring-wing systems, analogous to the quality factor of a linear spring-mass system. However, when spring-wing systems have internal structural damping, we observe that dynamic efficiency decreases with increasing $N$ on resonance, reducing the potential for useful energy storage and return. Overall, these results provide a generic framework to understand spring-wing systems which may enable us to learn more about the inter-relationships of morphology and actuation in flapping wing insects and birds.

\section{Acknowledgments}
NG and JL acknowledge support from the Mechanical and Aerospace Engineering department of UC San Diego. JG and SS acknowledge support from the US National Science Foundation CAREER grant 1554790 to S.S. and U.S. National Science Foundation Physics of Living Systems SAVI student research network (GT node
grant no. 1205878). Thanks to Dennis Wu for his invaluable work on the early prototypes of the robophysical flapper device and to Eugene Lin for his help fabricating silicone springs.

\onecolumngrid
\section*{References}
\twocolumngrid
\bibliography{references}{}

\bibliographystyle{abbrv}

\onecolumngrid
\newpage
\begin{center}
\textbf{\large Supplementary Material: Dimensional analysis of spring-wing systems reveals performance metrics for resonant flapping-wing flight} \\
\textsc{James Lynch, Jeffrey Gau, Simon Sponberg, Nick Gravish}
\end{center}

\section{Supplementary material}

\subsection{Computing added mass inertia}
\label{suppmat:addedmass}
The standard method for computing mean added mass is to treat the added mass as a cylindrical volume of fluid that surrounds the wing \cite{ellington_aerodynamics_1984-4}. The dimensions of the cylinder are defined by the dimensions of the wing: the radius is half the mean chord length, $\bar{c}/2$, the length is single wing span $R$ measured from wing hinge to wing tip, and the density is the fluid density $\rho$. The added rotational inertia is thus the rotational inertia of a cylinder of mass $m_A$ that rotates about its base: 

\begin{align*}
    m_A &= \rho \frac{\pi}{4} R \bar{c}^2 \\
    I_A &= \frac{1}{16} m_A \bar{c}^2 + \frac{1}{3} m_A R^2
\end{align*}

\noindent
In our system, with $\bar{c} =$ 3.5 cm, $R =$ 10 cm, $\rho_{H_2 O} =$ 977 $\frac{kg}{m^3}$, added mass inerita $I_A = 3.465\times 10^{-4} ~kg~m^2$.

\subsection{Structural damping modeling}
\label{suppmat:StructuralDamping}

Structural damping for generic oscillatory motion can be represented as an additional complex term in the spring stiffness parameter:
\begin{align}
K = k(1 + i\gamma)
\end{align}
However, this representation is not convenient for numerical modeling because of the complex term. If we assume that the oscillatory motion is sinusoidal, it is possible to express structural damping another way. Beginning with a generic spring-wing equation
\begin{align*}
    I_t\ddot{x} + k(1 + i\gamma)x + \Gamma\dot{x}^2 = 0
\end{align*}
\noindent
we make the substitutions $x = Xe^{i\omega t}$ and $\dot{x} = i\omega X e^{i \omega t} $:
\begin{align*}
     I_t\ddot{x} &+ k X e^{i\omega t} + \gamma k i X e^{i\omega t} + \Gamma\dot{x}^2=0
\end{align*}
\noindent
Using the definition $\frac{\dot{x}}{\omega} = i X e^{i\omega t}$, we can rearrange:

\begin{align}
    I_t\ddot{x} &+ k x + \frac{\gamma k}{\omega}\dot{x} + \Gamma\dot{x}^2=0
\end{align}

\noindent
Thus, the structural damping term can be represented as a viscous damper that is normalized by the oscillation frequency, implying frequency-independent viscous damping. 


\subsection{Derivation of the non-dimensional spring-wing equations}
\label{suppmat:nondim}

We introduce dimensionless time and angle parameters normalized to wing oscillation amplitude and frequency:

\begin{equation}
    \tau = \omega t, ~q_w = \frac{\theta}{\theta_0}, ~\dot{q}_w = \frac{\dot{\theta}}{\omega \theta_0}, ~\ddot{q}_w = \frac{\ddot{\theta}}{\omega^2 \theta_0} \nonumber
\end{equation}

\noindent
Plugging these terms into Eq. \ref{eqn:full_parallel} and rearranging coefficients, we obtain

\begin{flalign}
I_{t} \theta_0 \omega^2 \ddot{q}_{w} + k_{p} \theta_0 q_{w} + \frac{\gamma k_{p}}{\omega} \omega \theta_0 q_w + \Gamma \theta_0^2
\omega^2  |\dot{q}_w|\dot{q}_w &= T(t) \nonumber \\ 
\nonumber \\
\ddot{q}_{w} + \frac{k_p}{I_t\omega^2} q_w +\frac{\gamma k_p}{I_t \omega^2} \dot{q}_{w} + \frac{\Gamma \theta_0}{I_t} |\dot{q}_w|\dot{q}_w &= \frac{T(t)}{I_t\theta_0 \omega^2 } \nonumber \\
\nonumber \\
\ddot{q}_w + \hat{K}_p q_w + \gamma \hat{K}_p \dot{q}_w + \frac{1}{N} |\dot{q}_{w}|\dot{q}_{w} &= \hat{T}_p(t)
\label{eqn:parallel_nondimSUPP}
\end{flalign}

Eq. \ref{eqn:parallel_nondimSUPP} is a forced nonlinear oscillator defined by non-dimensional parameters $\hat{K}$, the reduced stiffness; $\gamma$, the structural damping loss modulus; and $N$, the Weis-Fogh number.

Performing a similar substitution for the series system we arrive at the equation
\begin{align}
    \ddot{q}_w + \hat{K}_s q_w + \gamma \hat{K}_s \dot{q}_w + \frac{1}{N} |\dot{q}_{w}|\dot{q}_{w} = \hat{T}_s(t)
    \label{eqn:series_nondimSUPP}
\end{align}
\noindent
which is identical to Eq. \ref{eqn:parallel_nondimSUPP} except for the normalized torque, which is now defined as

\begin{equation}
\hat{T}_s(t) = \frac{\hat{K}_s}{\theta_0}\left(\phi(t) + \frac{\gamma}{\omega}\dot{\phi}\right)   
\end{equation}
\noindent


\subsection{Computing drag torque coefficient}
\label{suppmat:aerodynamic}
We follow the standard methods for blade-element analysis of quasi-steady flapping wings. The wing is broken into differential chord elements, each of which experiences a differential aerodynamic torque,
\begin{align}
\mathrm{d} Q_{aero} =\frac{1}{2} r \rho\left(r \dot{\theta}\right)^{2} C_{D}(\alpha) c(r) \mathrm{d}r
\end{align}
The differential torques along the wing can be integrated across the entire wing shape resulting in the following equations

For simplicity we express the velocity dependence of the aerodynamic torque as, $\dot{\theta}^2$, and the sign dependence on the direction of motion is implied. The aerodynamic torque is governed by both the wing speed and the aerodynamic torque constant, $\Gamma$, which itself is a function of wing geometry (wing radial length, $R$ and shape factor $r_3$), wing pitch angle ($\alpha$), and fluid density ($\rho$). The drag coefficient, $C_D(\alpha)$, is dependent on the pitch angle of the wing, $\alpha$, which is 0 when the wing is vertical, and $\pi/2$ when the wing is horizontal. From Dickinson \cite{dickinson_wing_1999} the drag coefficient at insect-relevant Reynolds numbers is estimated as

\begin{align}
C_D(\alpha) = 1.92 - 1.55 \cos(2.04 \alpha - 9.82)
\label{eqn:dickinsonCD}
\end{align}

\subsection{Design and fabrication of silicone springs}
\label{suppmat:springs}

We designed and 3D-printed two-piece molds for casting the springs. Each mold was treated with Ease Release 200$^{TM}$ (Smooth On) before being filled with a common silicone material used in soft-robotics, Dragonskin$^{TM}$ 30A silicone (Smooth-On) \cite{Marechal2020-op}. Care was taken to de-gas the silicone in a vacuum chamber before filling the mold. The silicone molds were allowed to cure in a positive pressure chamber for at least 24 hours before removal and use.

The dimensions of the springs were determined by the desired spring stiffness. The torsional stiffness of a silicone spring is given by the stiffness equation for a twisting cylinder:
\begin{equation}
k_s = \frac{\mu \pi R^4}{2L}
\end{equation}
where $\mu$ is the shear modulus, $R$ is the spring radius, and $L$ is the spring length. We used three spring designs of constant length and radius 13, 16, 18~mm corresponding to torsional stiffness values of $k_s$ = [0.163, 0.416,0.632]~Nm/rad. 

\subsection{Data Processing}
\label{suppmat:data_analysis}
Analysis of both experiment and simulation data relied on the wing and motor angle data. To generate a consistent sampling time of all experiments we interpolated position measurements to a constant sample rate. The measured angle data was filtered with a 5th-order Butterworth filter at cutoff frequency of 10Hz (approximately 2.5 times greater than the peak driving frequency). Velocity was computed through numerical differentiation of the filtered position, and similarly acceleration from the filtered velocity.

We observed that the wing trajectories were consistent with a single frequency sin wave except when actuation frequency or amplitude approached the epxerimental limits (at low-amplitude and high-frequency and at high-amplitude). We used a nonlinear least-squares sine fit to find amplitude and phase of the motor, $\phi(t)$, and wing, $\theta(t)$, trajectories respectively.


\subsection{Derivation of non-dimensional resonance frequency for series system}
\label{supp:bennettcalc}
The following derivation is based on the process in \cite{bennett_elastic_1987}.
For a system with a spring with stiffness $k$ in series with an actuator that drives a mass $m$ subject to aerodynamic loading, $\Gamma \dot{x}^2$, the power driving the mass, $P_w$ is the sum of the aerodynamic and inertial forces times the velocity:
\begin{equation}
    P_m = (F_a+F_i)\dot{x} = (\Gamma \dot{x}^2 + m \ddot{x})\dot{x}
\end{equation}
\noindent
The strain energy in the spring, $E$, is
\begin{equation}
    E = \frac{1}{2}k^{-1}(F_a+F_i) =  \frac{1}{2}k^{-1} \left(\Gamma \dot{x}^2 + m \ddot{x}\right)^2 
\end{equation}
\noindent
Since the motor must both move the mass and compress the spring, the actuator power, $P_{act}$, is defined

\begin{align}
    P_{act} &= P_m + \dot{E}\nonumber \\ 
    P_{act} &= \left(\Gamma \dot{x}^2 + m \ddot{x} \right) \left[\dot{x} + k^{-1}\left(2\Gamma \dot{x} \ddot {x} + m \dddot{x}\right)   \right]
    \label{eqn:actPower}
\end{align}

\noindent
If we assume that the mass follows a sinusoidal trajectory, $x(t) = x_0\sin{\omega t}$, we can compute the derivatives and plug into \ref{eqn:actPower}:

\begin{equation}
    P_{act} = x_0^2\omega^3 \cos{\omega t} \left[\Gamma x_0 \cos^2{\omega t} - m \sin{\omega t}\right]\left[1-k^{-1}\omega^2(2\Gamma x_0 \sin{\omega t} + m) \right]
\end{equation}

\noindent If we define a non-dimensional actuator power, $\hat{P}_{act} = \frac{P_{act}}{m x_0^2 \omega^3}$, we can get the non-dimensional expression:

\begin{equation}
    \hat{P}_{act} = \cos{\omega t}\left( N^{-1}\cos^2{\omega t} - \sin{\omega t} \right) \left[ 1 - \hat{K}^{-1}(2N^{-1}\sin{\omega t} +1) \right]
\end{equation}

\noindent
Bennett et al. showed that this actuator power expression is minimized over half an oscillation period when it is always greater than zero. The relationship between $\hat{K}$ and $N$ that guarantees that condition is

\begin{equation}
    \hat{K} = \sqrt{1+4N^{-2}}
    \label{eqn:optStiffSupp}
\end{equation}

\noindent
Eq. \ref{eqn:optStiffSupp} describes the of spring, mass, aerodynamic damping and oscillation amplitude to get resonant oscillation at a particular frequency. Recalling that $\hat{K} = \frac{k}{m \omega^2}$, $N = \frac{m}{\Gamma x_0}$, and natural frequency $\omega_n^2 = \frac{k}{m}$ we can get an expression for that frequency:

\begin{equation}
    \omega^2_r = \frac{k}{\sqrt{m^2+4\Gamma^2 x^2_0}}= \frac{\omega_n^2}{\sqrt{1+4N^{-2}}}
    \label{eqn:resFreqEqnSupp}
\end{equation}{}


\subsection{Derivation of non-dimensional wing torques in the parallel system}
\label{supp:parallel}

Here we provide the full derivation for the non-dimensional work presented in Eq. \ref{eqn:non_dim_parallel}. We start from the non-dimensional force terms in the parallel system dynamics (Eqn.~\ref{eqn:parallel_nondim}). We make the assumption of sinusoidal wing motion, such that
\begin{align*}
    q &= \sin(\tau) \\
    \dot{q} &= \cos(\tau) \\
    \ddot{q} &= -\sin(\tau) \\
    &= -q
\end{align*}
Substituting these expressions in for the individual force terms in the parallel system and dividing by the aerodynamic force in those equations results in
\begin{align*}
    \tilde{Q}_{aero} &= \cos^2(\tau) \nonumber \\
    \tilde{Q}_{inertial} &= -N q  \nonumber \\
    \tilde{Q}_{ideal elastic} &= \hat{K}N q  \nonumber \\
    \tilde{Q}_{structural} &= \gamma \hat{K} N \cos(\tau)  \nonumber \\
\end{align*}
In order to write $\tilde{Q}_{aero}$ and $\tilde{Q}_{structural}$ in terms of wing angle we can use the following trigonometric relationship
\begin{align}
    \cos^2(\tau) &= 1 - \sin^2(\tau) \nonumber \\
    & = 1 - q^2 \\
    \cos(\tau) &= \sqrt{1 - q^2}
\end{align}
Substituting in the expressions of $\cos(\tau)$ and $\cos^2(\tau)$ yields the non-dimensional work equations in terms of just the normalized wing angle, $q$.



\end{document}